\documentclass[a4paper,11pt]{article}
\usepackage{pos}

\usepackage{subcaption}
\usepackage[capitalize]{cleveref}
\usepackage{booktabs}
\usepackage{placeins}

\usepackage{booktabs, multirow} 
\usepackage{soul}

\usepackage{tabularx}
\usepackage[usestackEOL]{stackengine}

\title{Parameterization of the frequency spectrum of radio emission in the 30-80\,MHz band from inclined air showers}
\ShortTitle{Parameterization of the frequency spectrum}

\author*[a,b]{Sara Martinelli}
\author[a,b]{Felix Schlüter}
\author[a,c]{Tim Huege}
\affiliation[a]{Karlsruhe Institute of Technology (KIT), Institute for Astroparticle Physics, Karlsruhe, Germany}
\affiliation[b]{Universidad Nacional de San Martín (UNSAM), Instituto de Tecnologías en Detección y Astropartículas, Buenos Aires, Argentina}
\affiliation[c]{Vrije Universiteit Brussel, Astrophysical Institute, Brussels, Belgium}

\emailAdd{sara.martinelli@kit.edu}
\emailAdd{tim.huege@kit.edu}

\abstract{To exploit the wealth of information carried by the short transient radio pulses from air showers, the frequency spectra of the signals have to be investigated. Here, we study the spectral content of radio signals produced by inclined showers, with a focus on the 30-80\,MHz frequency band, as measured, for example, by the antennas of the Pierre Auger Observatory. Two exponential models are investigated and used to describe the spectral shape of the Geomagnetic and Charge-excess components of the pulses. The spectral fitting procedure of the models is described in detail. For both components, a parameterization of the frequency slope as a function of the lateral distance to the shower axis and the geometrical distance between core and shower maximum is derived. For the Geomagnetic component, a quadratic correction to the frequency slope is needed to better describe the spectrum, and it has been parameterized, too. These pieces of information can be employed in event reconstruction to constrain the geometry, in particular the core position. For the analysis, CoREAS simulations for showers covering energies from $10^{18.4}$\,eV up to $10^{20.2}$\,eV and zenith angles from $65.0\,^\circ$ up to $85\,^\circ$ have been used.}

\FullConference{%
  9th International Workshop on Acoustic and Radio EeV Neutrino Detection Activities - ARENA2022\\
  7-10 June 2022\\
  Santiago de Compostela, Spain}

\begin{document}
\maketitle

\section{Introduction}
\begin{figure}[b]
\centering
\includegraphics[width=0.76\linewidth]{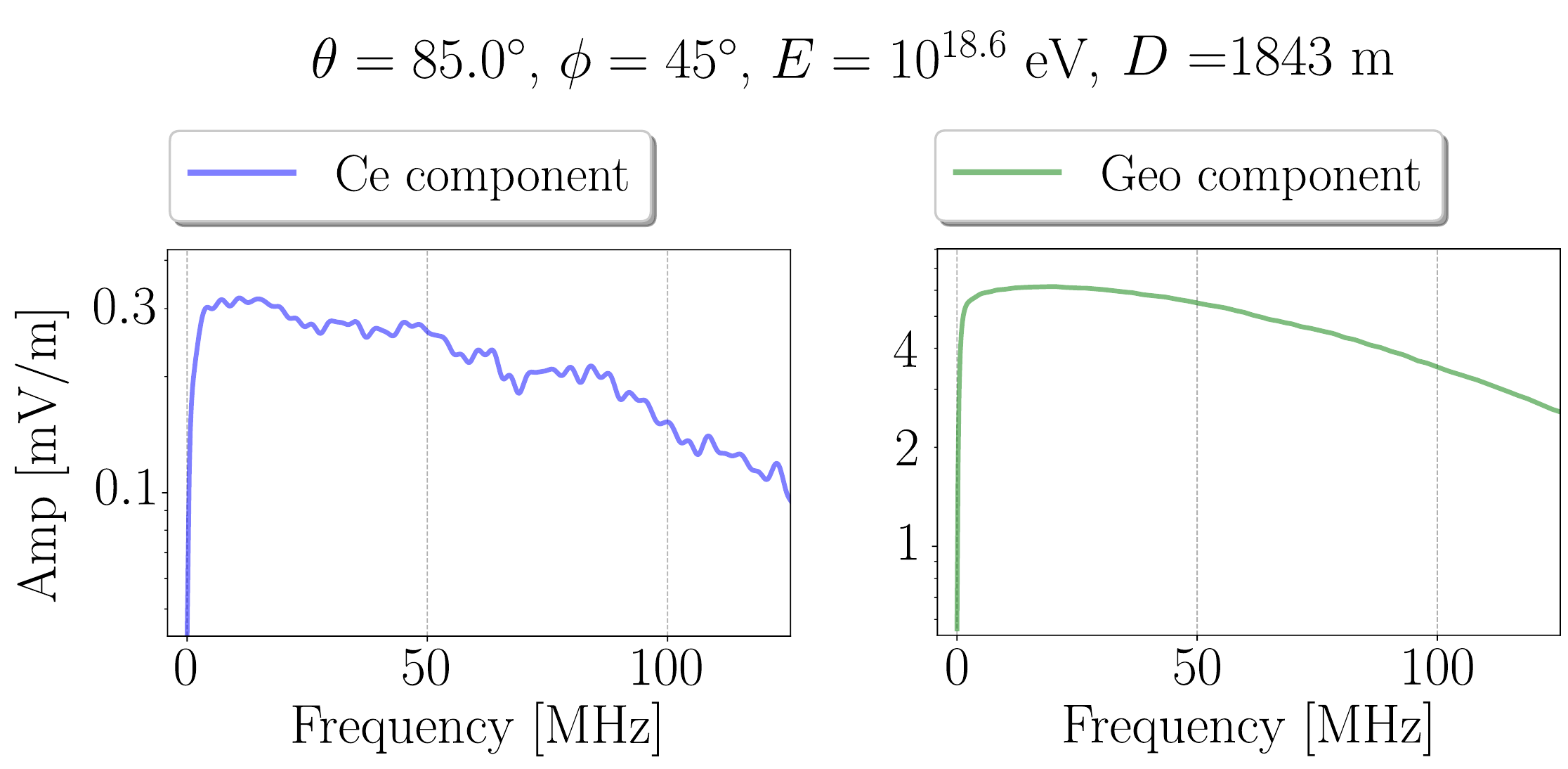}
\caption{Example of Charge-excess (left) and Geomagnetic (right) frequency spectra up to about 125\,MHz. The pulse is simulated at $D = 1843$\,m and belongs to a simulation characterized by $\theta = 85.0\,^\circ$, $\phi = 45.0\,^\circ$, $E = 10^{18.6}$\,eV. A logarithmic y-scale is adopted to highlight the curvature of the spectrum.}
\label{fig:spectrum_100}
\end{figure}
In a first approximation, the frequency spectrum of radio pulses produced by air-showers can be analytically described by an exponential function \citep{radio_emission_model, Welling}. Its shape depends on several factors, such as: the lateral distance $D$, the arrival direction, i.e. the zenith angle, the distance between core position and shower maximum $d_{\mathrm{max}}$, and the coherence of the emission. These dependencies manifest themselves as a change of the exponential function's slope. Previous works exploited the second-order dependence on the shower maximum to try and develop an $X_{\mathrm{max}}$ reconstruction-method \citep{Jansen_thesis, Canfora_thesis}.  A parameterization of the slope as a function of the lateral distance and the distance from the observer position to $X_{\mathrm{max}}$ can be found in \citep{Jansen_thesis}. The parameterization is valid for showers having zenith angles up to $60\,^\circ$ and the reconstruction method seems to be more suitable for single radio stations. Here, instead, we derive a parameterization of the Geomagnetic and Charge-excess spectral slopes valid for inclined showers. The parameterizations are expressed as a function of $d_{\mathrm{max}}$ and $r$, the lateral distance normalized to the Cherenkov radius of the shower. When adopting a logarithmic y-scale, the spectrum slightly diverges from a straight line (see figure \ref{fig:spectrum_100}). The curvature of the spectrum is described by introducing a quadratic correction to the exponential function's slope. In the following, a parameterization of the Geomagnetic spectral curvature is presented as well. Knowing the arrival direction, the antenna positions and $X_{\mathrm{max}}$, these pieces of information can be employed in event reconstruction to better constrain the geometry (e.g. the core position). The results of this work are valid for the frequency band of 30-80\,MHz, as exploited, for example, by the radio-antennas of AERA \citep{PierreAuger}, RD \citep{BjarniICRC2019}.  
\section{Simulation data-set}
For this analysis, we used a data-set containing 2158 simulated events generated with CORSIKA V7.7000 \citep{CORSIKA} and CoREAS \citep{CoREAS}. The proton-initiated inclined showers have been simulated adopting QGSJetII-04 \citep{QGSJET} and UrQMD \citep{UrQMD} as interaction models (high- and low-energy, respectively). The simulations have been produced using an optimized weight limitation method with a thinning level of $5\cdot 10^{-6}$ \citep{thinning}. The primary energies are distributed between $10^{18.4}$\,eV and $10^{20.2}$\,eV, with a logarithmic step of 0.2\,eV. The data set covers zenith angles $\theta$ from $65.0\,^\circ$ up to $85\,^\circ$, with a step of $2.5\,^\circ$, and 8 different azimuth angles $\phi$ = \{0, 45, ..., 315\}\,$^\circ$. Each simulation was performed using a star-shaped grid having 8 polar angles $\psi$ = \{0, 45, ..., 315\}\,$^\circ$, corresponding to 8 so-called arms in the shower reference frame. In order to reproduce the Pierre Auger Observatory \citep{PierreAuger} site conditions, the observation level is set to 1400\,m a.s.l.. The atmospheric model used is based on the monthly average atmospheric profile at Malargüe in October (model 27 available in CORSIKA \citep{October_profile}). The Geomagnetic field is set to 0.24\,Gauss. The refractive index at sea level $n_\mathrm{0}$ is given by $n_\mathrm{0}$ = 1 + 3.12$\cdot 10^{-4}$. 
\section{Method}
Given the $\vec{v}\times\vec{B}$ and $\vec{v}\times(\vec{v}\times\vec{B})$ components of the electric field, the Geomagnetic and Charge-excess electric fields \citep{Tim_review} are obtained from the equations \citep{Rad_en_release}: 
\begin{equation}
E_\mathrm{{\vec{v}\times\vec{B}}}\,(\vec{r},t) = E_{\mathrm{geo}}\,(\vec{r},t) + \cos\psi\cdot E_{\mathrm{ce}}\,(\vec{r},t),
\label{eq1}
\end{equation}
\begin{equation}
E_\mathrm{{\vec{v}\times(\vec{v}\times\vec{B})}}\,(\vec{r},t) = \sin\psi\cdot E_{\mathrm{ce}}\,(\vec{r},t).
\label{eq2}
\end{equation}  
The decoupling of equations \ref{eq1}, \ref{eq2} fails for polar angles $\psi = 0, 180\,^\circ$. Thus, pulses simulated on the $\vec{v}\times\vec{B}$-axis are excluded from the analysis. Thinning introduces noise in the simulations, which becomes more relevant the farther the observer position is from the shower axis \citep{Schluter:2022mhq}. For this reason, the analysis is restricted to pulses simulated at $r < 2$, where $r$ is given by:
\begin{equation}
r = D \,/\, r_{\mathrm{che}}.
\label{eq:distance}
\end{equation}
We calculate $D$ after shifting the core position by applying a correction-model for air-refractivity displacement \citep{refractive_displacement} and correcting for early-lateness effects of the observer positions \citep{el_effect}. Knowing the zenith angle, $X_{\mathrm{max}}$ of the shower and the observation level, we evaluate the vertical height above sea level of the shower maximum \textit{h}. The Cherenkov angle is calculated as $\theta_{\mathrm{che}} = \mathrm{arcos}\left(1/n(h)\right)$, where \textit{n(h)} is the refractive index at the height $h$. The latter is evaluated through the adopted CORSIKA atmospheric model, given the refractive index at sea level $n_{\mathrm{0}}$. We finally obtain the Cherenkov radius of the shower $r_{\mathrm{che}}$ as:
\begin{equation}
r_{\mathrm{che}} = \tan(\theta_{\mathrm{che}}) \cdot d_{\mathrm{max}}.
\end{equation}
\subsection{Spectral fitting}
To fit the spectra of Geomagnetic and Charge-excess components, we compared two exponential models, restricting the fit frequency-range to 30-80\,MHz. The first model considered is given by:
\begin{equation}
L(f) = A\cdot 10\,^{m_\mathrm{f}\cdot(f-f_\mathrm{0})},
\label{eq:lin}
\end{equation}
where \textit{A} is the spectral amplitude at a fixed frequency offset $f_\mathrm{0}$. The \textbf{\textit{frequency slope}} \textit{$m_\mathrm{f}$} represents the slope of a straight line describing the spectrum in log-space. Since $m_\mathrm{f}$ has a linear dependence on the frequency, the model \textit{L} is referred to as \textbf{\textit{linear model}}. In the fitting procedure of the linear model, \textit{A} and \textit{$m_\mathrm{f}$} are evaluated simultaneously. In the fit we chose as starting value for the slope \textit{$m_\mathrm{f}^*$} = -0.001\,$\mathrm{MHz}^{-1}$ and \textit{$A^*$} = $\mathrm{Amp}\,(f_\mathrm{0})$ for \textit{A}, where $\mathrm{Amp}\,(f_\mathrm{0})$ is the spectral amplitude of the frequency bin closest to $f_\mathrm{0}$. \\
The second model \textit{Q} is characterized by an additional parameter, as:
\begin{equation}
Q(f) = A\cdot 10\,^{m_\mathrm{f}\cdot(f-f_\mathrm{0}) + m_\mathrm{f2} \cdot (f-f_\mathrm{0})^2},
\label{eq:quad}
\end{equation}
where \textit{$m_\mathrm{f2}$} is the \textbf{\textit{quadratic correction}} to the frequency slope and allows for a description of the spectrum curvature. Since in log-space \textit{$m_\mathrm{f2}$} quadratically depends on the frequency, the model \textit{Q} is referred to as \textbf{\textit{quadratic model}}. To ensure the goodness and stability of the fit, we implemented a procedure consisting of three steps. The first step consists of fitting the linear model, as explained above. In the second step, the quadratic model is fitted, fixing the slope $m_\mathrm{f}$ to the value obtained in the first step, letting free only \textit{A} and \textit{$m_\mathrm{f2}$}. The starting value of \textit{A} is set to the value obtained by fitting the linear model, while the starting value of \textit{$m_\mathrm{f2}$} is chosen as \textit{$m_\mathrm{f2}^*$} = -0.1\,$10^{-5}\, \mathrm{MHz}^{-2}$. In a final step, all three parameters are optimized simultaneously, setting as starting values the values obtained in the second step.\\ 
In eqn. \ref{eq:lin} and \ref{eq:quad}, we introduced a frequency offset $f_\mathrm{0}$ to ensure that \textit{$m_\mathrm{f2}$} consists of a small correction to the amplitude and frequency slope of the linear model. We found that fixing $f_{\mathrm{0}}$ to 55\,MHz minimizes the difference between slopes $m_\mathrm{f}^\mathrm{lin}$ and $m_\mathrm{f}^\mathrm{quad}$ obtained by fitting, respectively, the linear and the quadratic models. We finally compared the models fitting the spectra of the entire data-set. The Geomagnetic component shows generally smaller $\chi^\mathrm{2}$-values and larger p-values when adopting the quadratic model, especially for $r < 0.5$ and $r > 1.5$.  
\subsection{Parameterization}
\begin{figure}[!b]
\centering
\includegraphics[width=0.76\linewidth]{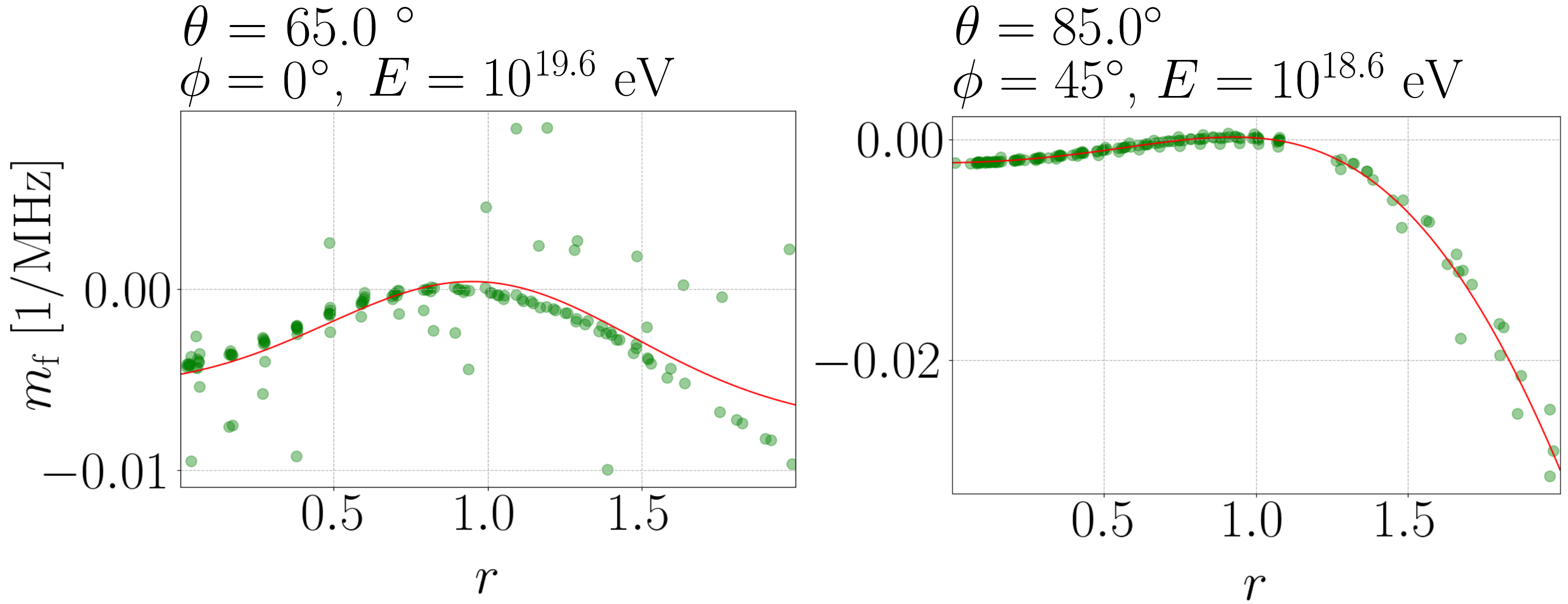}
\caption{Examples of lateral-distribution fitting of the parameter \textit{$m_\mathrm{f}$} from simulated showers characterized by $\theta = 65\,^\circ$, $E = 10^{19.6}$\,eV, $\phi = 0\,^\circ$ (left plot) and $\theta = 85\,^\circ$, $E = 10^{18.6}$\,eV, $\phi = 45\,^\circ$ (right plot). In the figure, we show the Geomagnetic slopes from the spectral fits (green points) as a function of $r$. The resulting curves obtained by fitting $f(r)$, keeping all five parameters free, are shown in red. }
\label{fig:geo_slope_free_ex}
\end{figure}
We express the Geomagnetic component in terms of frequency slope and quadratic correction, while for the Charge-excess component we adopt the linear model. The parameterization procedure consists of several stages. In the first stage, we fit the lateral distribution of the parameters \textit{$m_\mathrm{f}$}, \textit{$m_\mathrm{f2}$} for single events by means of the function:
\begin{equation}
f(r) = A_\mathrm{1} \cdot\left[ - e^{B \cdot (r-r_\mathrm{0})} + A_\mathrm{2}\cdot e^{-C \cdot (r-r_\mathrm{0})^2}\right],
\label{eq:param_function}
\end{equation}
keeping all five parameters $A_\mathrm{1}$, $A_\mathrm{2}$, $B$, $C$ and $r_{\mathrm{0}}$ free, as shown in the examples of figure \ref{fig:geo_slope_free_ex}. We express the parameter estimated in each individual free fit as a function of $d_{\mathrm{max}}$. In the following stages of the analysis, each parameter - one more in each step - is fixed through a parameterization as a function of $d_{\mathrm{max}}$. For example, after fixing the parameterization of $r_\mathrm{0}$, we repeat the fits with the remaining four free parameters. We look at the new parameters distributions expressed as a function of $d_{\mathrm{max}}$ and we fix the next parameterization. We fix the parameterization in the order of $r_\mathrm{0}$, \textit{B/C}, $A_\mathrm{2}$, $C$, $A_\mathrm{1}$ through fitting each distribution by the chosen function. The complexity of the fitting functions has been tuned until we considered the results adequate.  
\section{Results} 
\subsection{Geomagnetic slope}
\begin{figure}[h!]
\centering
\includegraphics[width=1.0\linewidth]{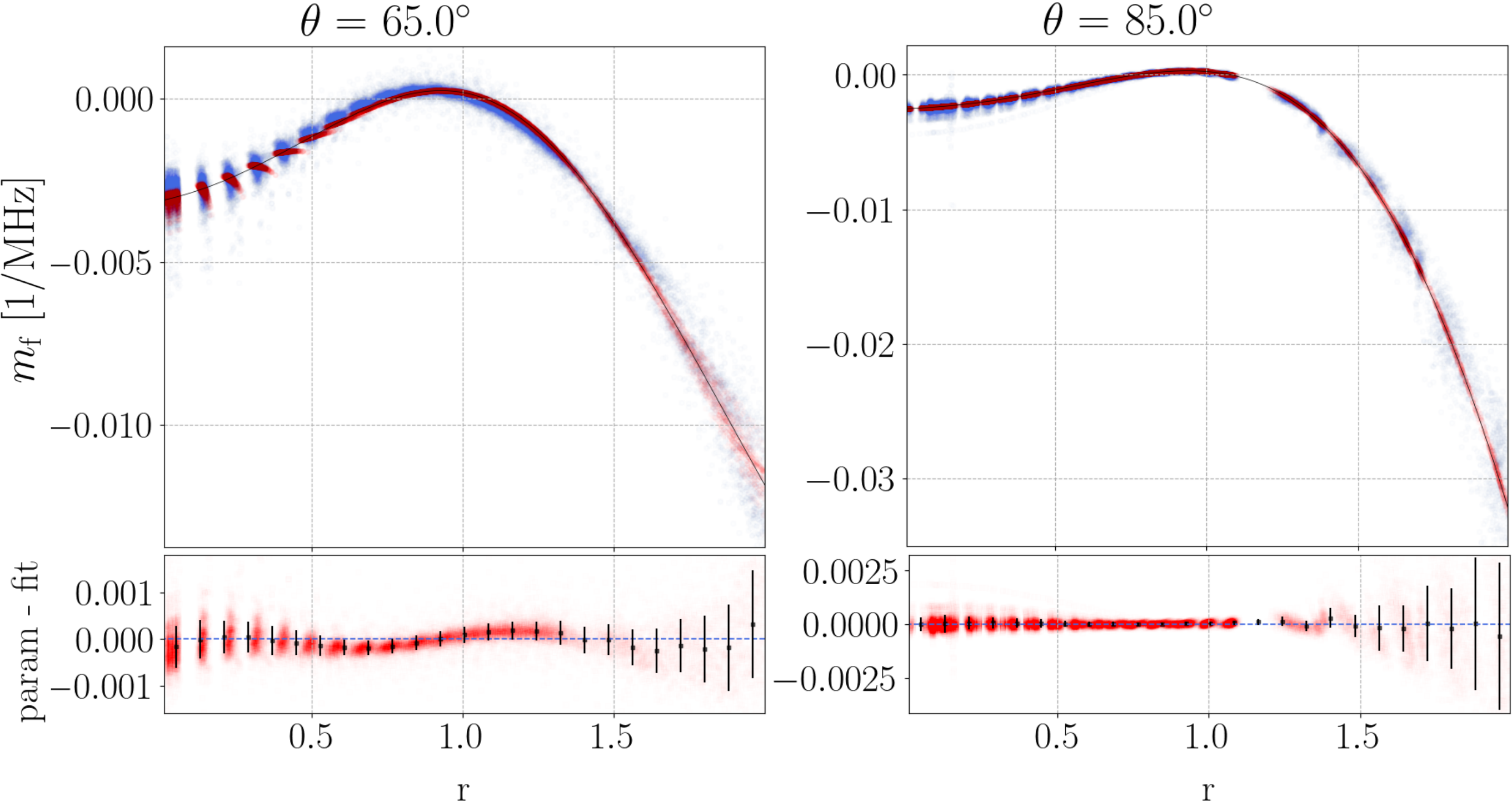}
\caption{Examples of Geomagnetic slope's lateral distribution and respective parameterized values. In the plots, we show values corresponding to different air-showers characterized by $\theta = 65\,^\circ$ (left), $85\,^\circ$ (right), as a function of the observer positions. In the upper plots, the slopes obtained by fitting the spectra (blue) are compared with the slopes calculated through the parameterization (red). The average parameterization curve (black) is illustrated, too. The lower plots show the slopes residuals (red) and the mean values (black) together with the standard deviations for each distance bin.  }
\label{fig:geo_slope_full_param_example}
\end{figure}

The functions used to fix the parameterizations of the Geomagnetic slope are listed below. 
\begin{equation}
r_\mathrm{0}\,(d_{\mathrm{max}}) = \frac{-3.558}{d_{\mathrm{max}}}\cdot \log(d_{\mathrm{max}})+1.738
\label{eq:r0_geo_slope}
\end{equation}
\begin{equation}
\frac{B(d_{\mathrm{max}})}{C\,(d_{\mathrm{max}})} = \frac{-7.078}{d_{\mathrm{max}}}\cdot \log(d_{\mathrm{max}})+1.625 
\label{eq:ratio_geo_slope}
\end{equation}
\begin{equation}
A_\mathrm{2}(d_{\mathrm{max}}) = \frac{3.468}{d_{\mathrm{max}}}\cdot \left[\log(0.2335 \cdot d_{\mathrm{max}})+0.4805\,\mathrm{km} \right]+0.5863 
\end{equation}
\begin{equation}
C\,(d_{\mathrm{max}}) = 0.9985^{- d_{\mathrm{max}}/\mathrm{km} } - \frac{0.2155}{\mathrm{km}}  \cdot \log(d_{\mathrm{max}} - 11.69\,
\mathrm{km} ) + 0.6492
\label{eq:C_geo_slope}
\end{equation}
\begin{equation}
A_\mathrm{1}(d_{\mathrm{max}}) = \left[\frac{-0.3792}{d_{\mathrm{max}}} \cdot \log(0.2008 \cdot d_{\mathrm{max}}) + \frac{4.057\cdot 10^{-5}}{\mathrm{km}}\cdot d_{\mathrm{max}}  + 0.0359\right]\,\mathrm{MHz}^{-1}
\end{equation} 
In the equations, $d_{\mathrm{max}}$ is expressed in units of km. In figure \ref{fig:geo_slope_full_param_example}, we show the average parameterization curve for subsets of simulations having $\theta = 65\,^\circ$ and $\theta = 85\,^\circ$, as examples. Here, we also compare the slopes obtained through spectral fitting with the values analytically calculated by exploiting the parameterization. As shown in the bottom plots, even though further improvements are surely possible in both cases, the slope residuals are generally small and distributed around zero. 
\subsection{Geomagnetic quadratic correction}
In comparison with the slope, for the parameterization of the quadratic term we had to pay more attention to the starting values and the fit bounds of each parameter. We also had to include fit uncertainties introducing an uncertainty model in the spectral-fitting procedure $\sigma(f)$ as:
\begin{equation}
\sigma\,(f) = a \cdot \mathrm{Amp}^\mathrm{max} + b \cdot \mathrm{Amp}(f).
\label{eq:error_estimation}
\end{equation}
fixing $a\,=\,b\,=\,0.1\cdot10^{-3}$. The two terms of eqn.\ref{eq:error_estimation} take into account the noise introduced by particle thinning in the simulations. The standard deviations obtained by fitting the spectra were adopted as fit uncertainties in the lateral fits. The functions used to fix the parameterizations can be found below. When comparing the spectral fitting values and the parameterized ones, the residuals are centered around zero and get smaller with increasing zenith angle.
\begin{equation}
r_\mathrm{0} (d_{\mathrm{max}}) =  1.219 + \frac{0.0019}{\mathrm{km}} \cdot d_{\mathrm{max}} + \frac{ 8.768\,\mathrm{km}}{d_{\mathrm{max}}^2}\cdot \left[\log(558448\cdot d_{\mathrm{max}} )- d_{\mathrm{max}}\right] 
\end{equation}
\begin{equation}
\frac{B(d_{\mathrm{max}})}{C(d_{\mathrm{max}})} = -\,1.014^{-d_{\mathrm{max}}/\mathrm{km}}\,-\, 1.0000095^{(d_{\mathrm{max}}/\mathrm{km})^2}+ 1.824 
\end{equation}
\begin{equation}
A_\mathrm{2}(d_{\mathrm{max}}) = -\,73.28 -\,\frac{0.000996}{\mathrm{km}} \cdot d_{\mathrm{max}} - \frac{74.21}{d_{\mathrm{max}}^2}\cdot\left[\log(0.0482\,\mathrm{km} \cdot d_{\mathrm{max}})\,-\,d_{\mathrm{max}}^2\right] ,
\end{equation}
\begin{equation}
C\,(d_{\mathrm{max}}) = \frac{59.77}{d_{\mathrm{max}}}\cdot \log(0.0898 \cdot d_{\mathrm{max}}) \,+\frac{0.0018}{\mathrm{km}} \cdot d_{\mathrm{max}} -0.2631
\end{equation}
\begin{equation}
A_\mathrm{1}(d_{\mathrm{max}}) = \left[\frac{1.169}{\mathrm{km}} \cdot 10^{-6} \cdot d_{\mathrm{max}} + 2.957 \cdot 10^{-5}\right]\,\mathrm{MHz}^{-2}
\end{equation}
\subsection{Charge-excess slope}
\begin{figure}[!t]
\centering
\includegraphics[width=1.0\linewidth]{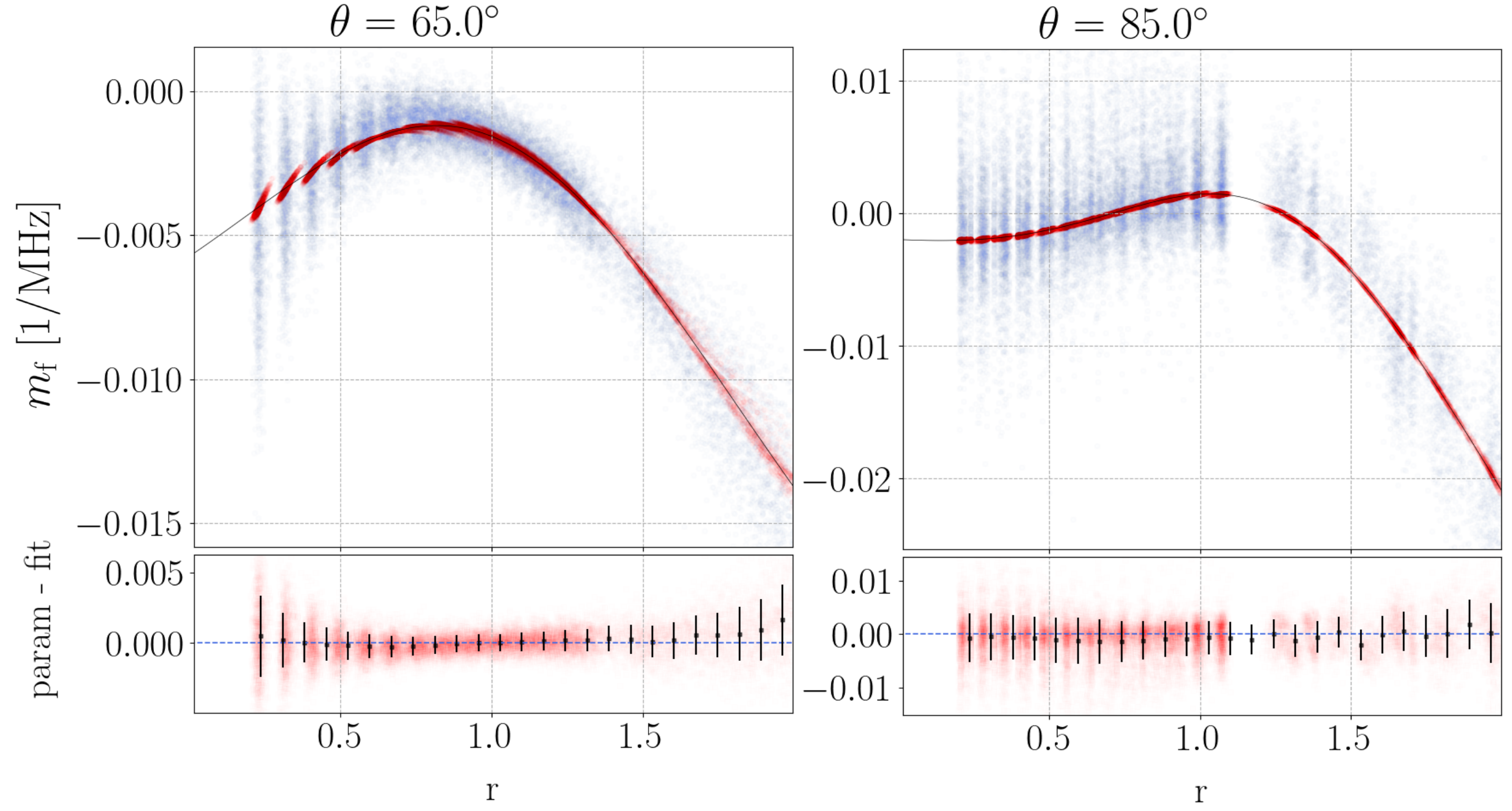}
\caption{Examples of Charge-excess slope's parameterization for simulations having $\theta = 65\,^\circ$ (left) and $\theta = 85\,^\circ$ (right). In the bottom plots, we show the slope residuals, as done in figure \ref{fig:geo_slope_full_param_example} (the same color code of the figure is adopted, too). }
\label{fig:ce_full_param_example}
\end{figure}

As for the Geomagnetic quadratic correction, the standard deviations obtained by fitting the spectra were adopted as fit uncertainties. The spectral fits were executed using the error-estimation model detailed in eqn.\ \ref{eq:error_estimation}, setting $a\,=\,0.25$ and $b\,=\,0.20$ . In order to guarantee the stability of the fit, starting values and fit bounds for all the parameters of $f(r)$ were set as well. Additionally, all the signals simulated at $r < 0.2$ were excluded from the analysis. Below, we report the functions used to fix the parameterizations. In figure \ref{fig:ce_full_param_example},the parameterization obtained for simulations having $\theta = 65\,^\circ$ and $\theta = 85\,^\circ$ are shown as examples. 
\begin{equation}
r_\mathrm{0} (d_{\mathrm{max}}) =- \frac{2.021}{d_{\mathrm{max}}} \cdot \log(d_{\mathrm{max}}) + 1.320 
\end{equation}
\begin{equation}
\frac{B(d_{\mathrm{max}})}{C(d_{\mathrm{max}})} = \frac{0.0016}{\mathrm{km}} \cdot d_{\mathrm{max}} + 0.2764
\end{equation}
\begin{equation}
A_\mathrm{2}(d_{\mathrm{max}}) = \frac{0.00085}{\mathrm{km}} \cdot d_{\mathrm{max}} + 0.8870 
\end{equation}
\begin{equation}
C\,(d_{\mathrm{max}}) = -\frac{10.54}{d_{\mathrm{max}}} \cdot \log(d_{\mathrm{max}}) + 2.570  
\end{equation}
\begin{equation}
A_\mathrm{1}(d_{\mathrm{max}}) = \left[\frac{0.0270}{d_{\mathrm{max}}} \cdot \left[ \log(388  \cdot d_{\mathrm{max}}) - 4.748\,\mathrm{km} \right]+0.0089 \right]\, \mathrm{MHz}^{-1}
\end{equation}
\section{Conclusion}
We presented a study on the spectral content of the Geomagnetic and Charge-excess emissions from inclined showers in the interval 30-80\,MHz. We found that a quadratic correction to the frequency slope is needed to better describe the Geomagnetic spectrum. A parameterization of the frequency slope as a function of the lateral distance and $d_{\mathrm{max}}$ is now available for both components and can be employed to constrain the geometry in reconstruction algorithms. The parameterization of the Geomagnetic quadratic correction here presented can be exploited, too.  
\\\\\\
\bibliographystyle{JHEP}
\setlength{\bibsep}{0pt plus 0.3ex}
\bibliography{bibliography}

\end{document}